\documentstyle[graphicx]{aipproc}

\begin{document}

\title{First Principles Studies of KNbO$_3$,
KTaO$_3$ and LiTaO$_3$ Solid Solutions}

\author{Serguei Prosandeev, Eric Cockayne and Benjamin Burton}

\address{Ceramics Division, Materials Science and Engineering
Laboratory, National Institute of Standards and Technology,
Gaithersburg, MD 20899-8520}

\maketitle

\begin{abstract}

 KTaO$_3$-based solid solutions exhibit a variety of
interesting physical phenomena.  To better understand these
phenomena, we performed first-principles calculations on
[K$_{1-x}$,Li$_{x}$]TaO$_{3}$ (KLT) and
K[Ta$_{1-x}$,Nb$_{x}$]O$_{3}$ (KTN) supercells.  Our results show
Li displacements and potential barrier heights in KLT that are in
excellent agreement with values obtained from experimental fits.
Dramatic changes in B-site dynamical charges occur in KTN in
response to changes in near neighbor (nn) coordination. These
effects can be explained by heterogeneity in the local electronic
dielectric permittivity.

\end{abstract}

\section*{Introduction}

 KTaO$_3$ (KT) is one of the rare perovskite-type materials that
remains cubic down to zero Kelvin. It is also interesting because
it is a quantum paraelectric and, therefore, its ferroelectric
phase transition at low temperatures is suppressed by zero-point
quantum vibrations.  KNbO$_3$ (KN) is ferroelectric even at room
temperature and exhibits a series of ferroelectric phase transitions
with decreasing temperature. A variety of interesting phenomena
occur when KT or/and KN form solid solutions via isovalent
substitution of A or B cations; [K$_{1-x},$Li$_x$]TaO$_3$ (KLT)
and K[Ta$_{1-x}$,Nb$_{x}$]O$_{3}$ (KTN) are solid solutions
which exhibit both order-disorder and soft-mode characteristics.
In this work, we use first-principles supercell calculations to
investigate lattice dynamics, ground-state structures, and
dynamical effective charges in KLT and KTN.

 Recent developments in quantum theory of polarization
(see the review in \cite{Resta}) allow one to make first-principle
(FP) computations of polarization and its derivative, dynamical
charge. Such computations show that dynamical charges are
anomalously large in ferroelectric perovskite-type crystals (see
e.g. Ref. \cite{Ghosez}), and this has a significant effect on the
soft-mode behaviors of these materials. Variations of dynamical
charges as functions of local chemical coordination, as occurs in
solid solutions, have not been extensively studied.  We present
systematic results on the dynamical charges as functions of
chemical configurations in KTN. The same tendencies were found for
PMN supercells.

\section*{Methods}

We use the Vienna {\it ab initio} simulation package (VASP)
\cite{Kresse1,Kresse2}, which treats electronic structure within
the framework of density function theory. Detailed computations
were performed for 40 ion supercells and a limited number of
computations were done for an 80 ion supercell. The basis vectors
were [002], [020] and [200] for the 40 atom cell (KLT40), and
[220], [022] and [202] for the 80 atom cell. The k-vector mesh was
constructed from 4x4x4 Monkhorst-Pack grid. Verification of this
computational scheme was performed for pure KTaO$_{3}$; the
self-consistent lattice constant $a$ = 3.96 \AA~ agrees with
previous LDA and GGA computations \cite{Singh}. The calculated
lattice constant is smaller than the experimental value $a$ =
3.983 \AA~ \cite{exp_KTO_a}.

 VASP computes interatomic forces and total energies for crystals,
and allows global relaxation as well as constrained relaxation of
chosen internal coordinates and/or lattice parameters. We used
frozen-phonon methods to obtain force constants for computing
lattice dynamics. To compute dynamical charges, we used Berry
phase analyses as implemented in the VASP code by Martijn Marsman.

\section*{KLT results}

We calculated the force constants matrix in KLT40 containing one
Li per cell. The reference structure for the force constant
calculations was obtained by first placing all ions on ideal
perovskite positions and then optimizing lattice parameters and
internal coordinates while maintaining full cubic symmetry. The
resultant cell had $a = 2 \times 3.956$ \AA.

 From the force constants, we obtained the dynamical matrix,
and the normal mode frequencies and eigenvectors.  Symmetry
analysis of KLT40 shows 14 $\Gamma_{15}$ triplets,
one which is acoustic and 13 which represent polar
transverse optical (TO) modes. In Table~\ref{kltphon.tbl},
we give the TO normal mode frequencies.

\begin{table}[h]

 \caption{Computed TO normal mode frequencies $\nu$
(in cm$^{-1}$) for cubic K$_7$LiTa$_8$O$_{24}$ (KLT).  Also shown
are relative amplitudes of each mode in the fully relaxed
tetragonal KLT structure $a_{rel}$ and relative contributions of
each mode to the tetragonal phase polarization $P_{rel}$ }

\begin{tabular}{ccccccccc}
$\nu$ & $a_{rel}$ & $P_{rel} $ & $\nu$  & $a_{rel}$ & $P_{rel}$ &
$\nu$  & $a_{rel}$ & $P_{rel}$ \\ \hline
 191 $i$ &  1.0000 &  0.7225 & 200 & 0.0034 &  0.0023 & 433 & 0.0095 & 0.0005 \\
 121     &  0.1215 &  0.2330 & 205 & 0.0864 &  0.0384 & 555 & 0.0093 & 0.0129 \\
 164     &  0.0217 &  0.0014 & 244 & 0.0358 &  0.0080 & 862 & 0.0030 & 0.0003 \\
 177     &  0.0017 & -0.0001 & 337 & 0.0283 &  0.0023 &     &   &   \\
 187     &  0.0806 & -0.0212 & 366 & 0.0126 & -0.0002 &     &   &   \\
\hline
\end{tabular} \label{kltphon.tbl} \end{table}

 For comparison, we also calculated the force constants
of a 40-atom cell of pure KT at the same lattice parameter. The TO
normal mode results are shown in Table~\ref{ktphon.tbl}. The force
constants in the two cases are nearly identical. The only
interatomic force constants which change by more than 0.41
eV/\AA$^2$ are those involving the 12 O ions nearest the Li site,
and which favor Li moving against its nearest neighor O sites.  In
fact, the one lattice instability found in KLT is essentially Li
motion (85\% of the dynamical matrix eigenvector) opposite its
nearest neighbor O (12\% of the dynamical matrix eigenvector).
Because KLT40 and KT have such similar force constant matrices, it
is not surprising that the normal mode spectra are very similar:
note that each LO mode in KT has a corresponding KLT40 mode with
very similar frequency. Thus, the FP results confirm a picture of
KLT in which Li are displaced off-center.  We emphasize, however,
that a quantitative description of the energetics and dynamics of
KLT requires that one also take into account the coordinated
motion of the ions surrounding the Li.

\begin{table}[h] \caption{Computed TO normal mode frequencies
$\nu$ (in cm$^{-1}$) and dynamical matrix eigenvectors for KTaO$_3$
at $a$ = 3.956~\AA.}
\begin{tabular}{ccccc}
$\nu$ & $u_{K}$ & $u_{Ta}$ & $u_{O\parallel}$ & $u_{O\perp}$
\\\hline\\
115 & --0.2989 &   0.5419 & --0.3937 & --0.4807 \\
205  &  0.8745 & --0.1737 & --0.2401 & --0.2715 \\
555  &  0.0012 &   0.0341 & --0.8531 &   0.3682 \\
\hline
\end{tabular} \label{ktphon.tbl} \end{table}

  The largest reduction in energy in KLT40 occurs when
Li is displaced along a [001] vector. Our results thus support a
picture of KLT in which each Li is displaced in one of the 6
[001]-type directions. We performed a full ionic relaxation of
KLT, polarized along $\hat{z}$. A breakdown of the displacement
pattern into normal mode coordinates is shown in
Table~\ref{kltphon.tbl}.  Although there is some mixture of higher
frequency modes due to anharmonic coupling, the displacement
pattern is dominated by the eigenvector of the lattice
instability.  A more interesting picture emerges when we
investigate the contribution of each mode to the polarization of
the relaxed structure.   Because the 121 cm$^{-1}$ mode
corresponding to the soft mode of pure KT has such a large
mode effective charge, it contributes 23\% of the total
polarization of KLT40.

We investigated Li energy as a function of displacement direction
and found that the smallest barrier between [001] wells are saddle
points in the [110] directions. Fig. 1 shows the energy as a
function of Li displacement $d$~ in the [001] and [110] directions
(see also \cite{Postnikov}). Since the barrier energy is
comparable with the thermal energy, Li can jump between
neighboring [0,0,$\overline{d}$] wells via [110] saddle points.

If all ions except for the Li are fixed, the Li displacements and
potential barriers are very small relative those deduced
experimentally (by fitting a mean field expansion to experimental
data \cite{Huchli}) (upper curves in Fig. 1, see also
\cite{Postnikov}). However, fully relaxed calculations (lower
curves in Fig. 1) yield good agreement: the equilibrium Li
displacement is [0,0,0.99 \AA] with $a$ = 3.96 \AA~ and [0,0,1.07
\AA] with $a$ = 4.0 \AA; the corresponding adiabatic potential
barriers are $\sim$103 meV (1190 K) and $\sim$117 meV (1360 K),
respectively. The calculated values are somewhat larger than the
experimentally estimated barrier of 86 meV (998 K) \cite{Huchli},
but they agree rather well with the 1200 K barrier reported by
\cite{Toulouse}. Computed barrier heights may be systematically
larger than fitted values if the latter are reduced by
Li-tunneling between neighboring wells over excited states.  For
the 80 atom supercell we obtained the Li displacement 1.01 \AA~
and the energy barrier $U = $118 meV (1369 K), which is also
somewhat larger than the experimental value of 86 meV (998 K)
\cite{Huchli}.

Fig. 1(b)  represents our computations of Li potential relief
along a vector connecting neighboring Li-wells, e.g. [0,0,d] and
[0,d,0] via a minimum energy [011] saddle.  We computed the
``frozen coordinate" ({\it i.e} all ions frozen except for Li) energies, as
well as the adiabatic energies ({\it i.e} with relaxed coordinates
of the non-Li ions). Clearly, ionic relaxation promotes Li-hopping by
reducing potential barriers.

\begin{figure}
\resizebox{0.9\textwidth}{!}{ \includegraphics{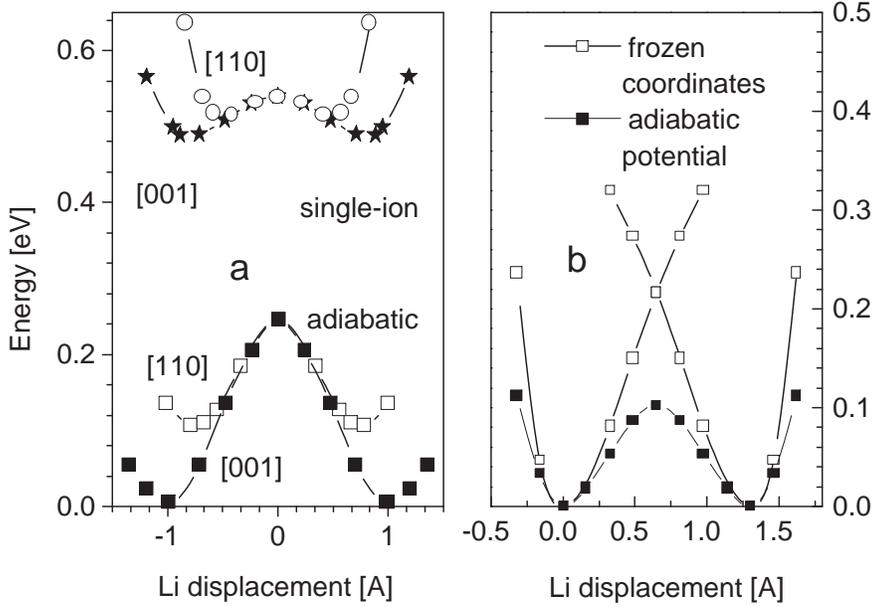}}
\caption{Li adiabatic and single-ion potential relief in KTaO$_3$
in [001], [110] directions (a) and in the direction between the Li
wells (b) } \label{Fig1}
\end{figure}

In a relaxed supercell, cations displace in the same direction as
Li, and anions displace in the opposite direction. This result
supports the idea of coupling between Li displacements and
ferroelectrically active optical host ion modes \cite{Vugmeister}.
The largest local deformation is exhibited by the four oxygens
that are closest to Li. For example, in the 80 atom cell where the
Li-displacement is [0,0,1.01 \AA], their displacements are
[0,-0.074 \AA,-0.074 \AA].
In the 40 ion supercell these displacements are similar, [0,-0.093
\AA,-0.093 \AA]. Such distortion was proposed \cite{Glinchuk} to
explain experimental data on the temperature dependence of
photocurrent in [K$_{1-x}$,Li$_{x}$]TaO$_{3}$~ solid solutions.

These results are in excellent agreement with computations for
LiTaO$_{3}$~ and LiNbO$_{3}$~ \cite{Cohen}, which also indicated a
large oxygen displacements similar to those in our calculations.
These oxygen displacments were explained in terms of a reduction in
elastic energy that occurs because oxygens prefer to rotate about
B site cations rather than directly approaching them \cite{Cohen}.

Based on the fact that Li hop over a saddle point, we derived the
following expression describing the temperature dependence of the
relaxation time:

\begin{equation}
\label{eqtau} \tau = A\int\limits_{ - \infty }^\infty {e^{\left(
{U + ay^2} \right) / k_B T}dy} = A\sqrt {\frac{\pi k_B T}{a}}
e^{U\left( T \right) / k_B T} = \tau _0 \left( T \right)e^{U\left(
T \right) / k_B T}
\end{equation}

\noindent where: $\tau$~ is the relaxation time; $k_B$~ is the
Boltzmann constant; $a$~ is a constant. The temperature dependence
of the prefactor is a consequence of the existense of different
paths when Li overcomes the potential barrier. The temperature
dependence of the average barrier, $U(T)$, is governed by the
dynamics of the ions surrounding a Li. This dependence can be
understood if one introduces spatial and thermal potential barrier
fluctuations

\begin{equation}
\label{eqU}
\begin{array}{c}
 e^{ - U(T) / k_B T} = \frac{\alpha \beta }{\pi k_B T}\int\limits_{ - \infty
}^\infty dx {\int\limits_{ - \infty }^\infty{dy~e^{\left( {U_0 + x
+ y} \right) / k_B T}e^{ - \alpha ^2x^2 - \beta ^2\left( {y / k_B
T} \right)^2}}
 } \\
 = e^{ - U_0 / k_B T + 1 / 4\beta ^2 + 1 / 4\alpha ^2k_B^2 T^2} \\
\end{array}
\end{equation}

\noindent From comparison of the left and right sides of this
equality one finds $U\left( T \right) = U_0 - k_B T / 4\beta ^2 -
1 / 4\alpha ^2k_B T$. At high temperature the potential barrier
decreases linearly with temperature because of thermal
fluctuations (in excellent agreement with experiments performed
for [K$_{0.957}$,Li$_{0.043}$]TaO$_{3}$ \cite{Trepakov}). At low
temperatures the barrier height decreases because of spatial
fluctuations. Also, critical slowing down of relaxation is common,
at low temperature, but in KLT, Li-tunneling is a competitive
process \cite{Kleemann}.

\begin{figure}
\begin{center}
\resizebox{0.8\textwidth}{!}{ \includegraphics{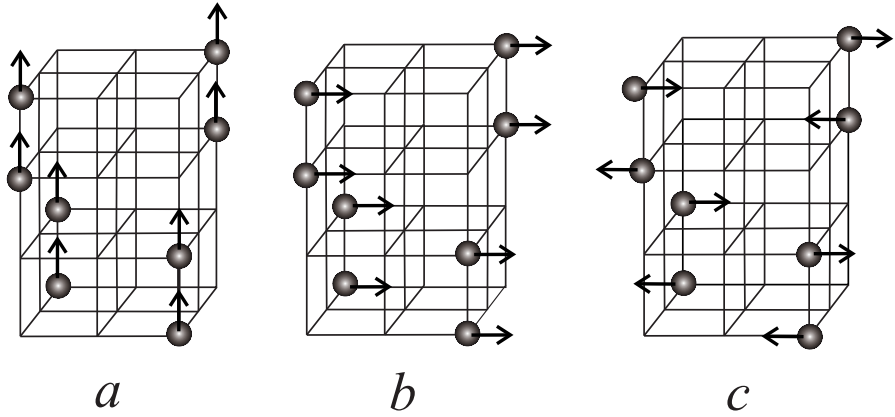}}
\caption{A-site configurations in a
[K$_{14/16}$Li$_{2/16}$]TaO$_3$ supercell (undecorated sites are
occupied by K): a) $(zz)$, b) $(xx)$, c) $(x\overline{x})$ }
\label{Li2}
\end{center}
\end{figure}

We also performed computations for 80 atom supercells with Li-Li
pairs (Fig. \ref{Li2}). A Li-Li pair was introduced into an 80-ion
supercell such that nn A-sites (separation vector $a \hat{z}$)
were occupied by Li. We considered the following Li-displacement
configurations for this pair: $(zz)$, $(xx)$~ and
$(x\overline{x})$. The energy required to convert a $(zz)$~ to
$(x\overline{x})$~ configuration is 0.27 eV (2990 K) in excellent
agreement with the reported barrier height of 2800 K
\cite{Toulouse,Trepakov}, attributed to the rearrangement of Li
pairs \cite{Dousenau}. Significantly, the excitation energy
required to move Li pairs into the $(xx)$~ configuration, 0.26 eV,
is close to the energy required to excite the $(x\overline{x})$
configuration. This occurs because of the indirect dipole-dipole
interaction between Li ions in Li-Li pairs over the KT-like soft
mode.

The computed dynamical charge of Li in KTaO$_{3}$:Li is 1.2 e at
the equilibrium position but it becomes lower than 1.0 at small Li
displacements.

\section*{KT and KN results}

\begin{figure}
\begin{center}\resizebox{0.7\textwidth}{!}{\includegraphics{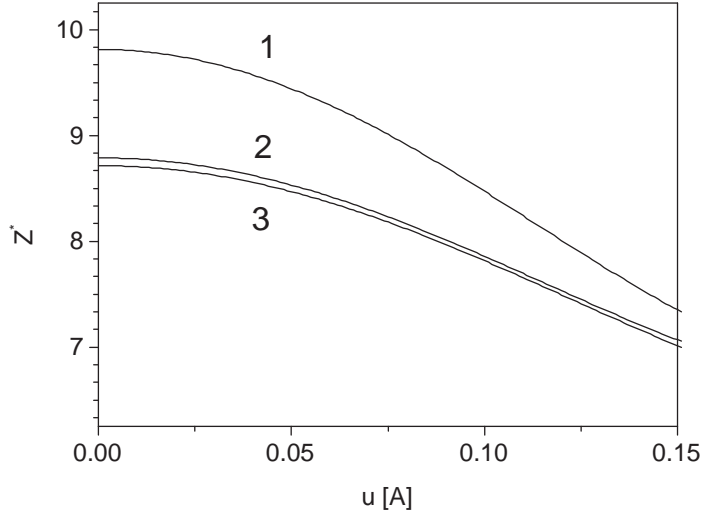}}
\caption{Dependence of the dynamical charge on the cation
displacement: 1.~KN, $a=4.015$\AA. 2.~KT $a=4.015$\AA. 3. KT
$a=3.98$\AA.} \label{Fig-pot}
\end{center}
\end{figure}

The dynamical charges of Ta ($Z^{\ast}_{Ta}$) and Nb
($Z^{\ast}_{Nb}$), in pure KT and KN, respectively, were
calculated as functions of [001] displacement in cubic structures
with $a$ = 4.015 \AA, using a 6x6x6 Monkhorst-Pack k-point grid.
The results are shown in Fig. \ref{Fig-pot}.

We also computed electrostatic potentials on the ions and their
derivatives.  Displacement induced reductions in the second
derivative of the potential were consistent with changes in
$Z^{\ast}_{Ta}$~ and $Z^{\ast}_{Nb}$. Indeed one can write a
general relation: $E_i(u)={\xi_{i}(u)}Z^{*}_{i}u_{i}$~ where
$\xi_{i}(0)\sim1/\varepsilon_{\infty}$~ and
$\varepsilon_{\infty}$~ is a high frequency dielectric constant.
Hence the reduction in local field (and charge-transfer
\cite{Resta}) that is caused by ionic displacement is responsible
for the calculated decrease in dynamical charge. Note that this
decrease is rather large: e.g. for Nb in KN $Z^{\ast}_{Nb}$ = 9.8
at u=0; 8.5 at u=0.1 \AA; and 6.8 at u=0.2 \AA.

\section*{KTN results}

Results of $Z^{\ast}$~ computations for KTN supercells are listed
in Table 3 and some results for [001] ordered supercells are shown
in Fig. 4. Calculations for different [hkl] ordered
superstructures, (where [hkl] indicates an ordered sequence of
layers perpendicular to the [hkl] vector) were all done with an
8x8x8 Monkhorst-Pack grid. All computations were performed for the
\textit{cubic structure} with the fixed lattice parameter $a=3.96$
\AA. This was done in order to show the pure effect of the
neighbors on the dynamical charge. The effect of the ionic
displacements were discussed in the previous section.

\begin{table}[htbp]
\caption{Dynamical charges in KTN supercells: the first two
columns represent the relative numbers of the Nb, $x_{Nb}$, and
Ta, $x_{Ta}$, planes; the vector [\textbf{hkl}] shows the
direction perpendicular to the planes.}
\begin{tabular}
{p{45pt}p{45pt}p{45pt}p{45pt}p{45pt}p{45pt}p{45pt}}
\textbf{x$_{Nb}$}& \textbf{x$_{Ta}$}& \textbf{[hkl]}&
\textbf{${Z^\ast}_{Nb,zz}$} & \textbf{${Z^\ast}_{Nb,xx}$} &
\textbf{${Z^\ast}_{Ta,zz}$} & \textbf{${Z^\ast}_{Ta,xx}$} \\
\hline \\ 1& 0& & 9.91& 9.91& &
  \\
0& 1& & & & 8.79& 8.79 \\ 1/2& 1/2& [001]& 9.06& 9.88& 9.56& 8.81
\\
1/3& 2/3& [001]& 8.78& 9.88& 9.30 \par 9.31& 8.80 \par 8.81 \\
1/4& 3/4& [001]& 8.62& 9.88& 9.18 \par 9.23 \par 9.16& 8.80 \par
8.80 \par 8.80 \\ 2/3& 1/3& [001]& 9.31 \par 9.31& 9.89 \par 9.89&
9.89& 8.82 \\ 3/4& 1/4& [001]& 9.47 \par 9.41 \par 9.47& 9.89 \par
9.90
\par 9.89& 10.06&
8.81 \\ 1/2& 1/2& [111]& 8.75& 8.75& 9.22& 9.22 \\ 1/2& 1/2&
[110]& 9.87& 9.23& 8.84& 9.68 \\
\end{tabular}
\label{tab1}
\end{table}

\begin{figure}[tbp]
\resizebox{0.9\textwidth}{!}{\includegraphics{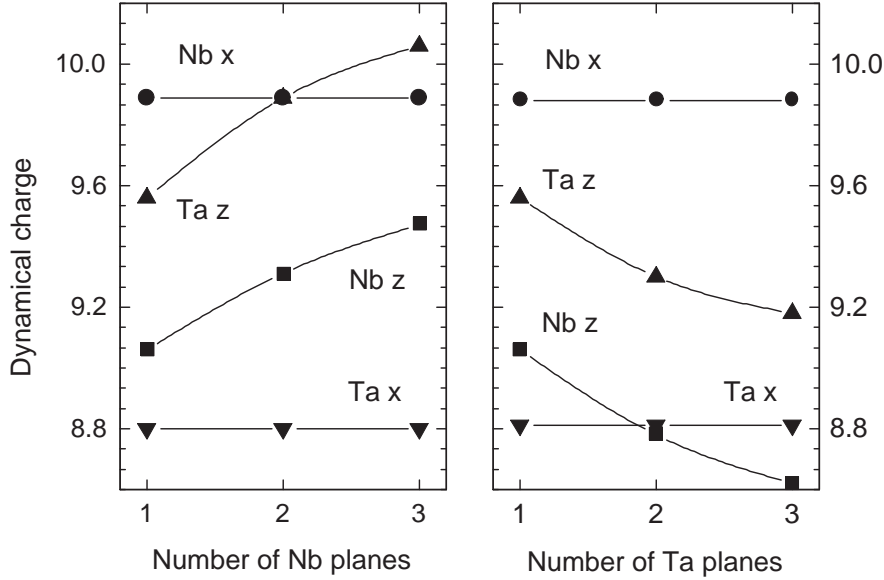}}
\caption{Dependence of the dynamical charge on the number of Ta-
and Nb-  planes in KTN (001) supercells} \label{Fig3}
\end{figure}

In the cubic KN, $Z^{\ast}_{Nb,zz} = 9.9$, which is larger than
$Z^{\ast}_{Ta,zz} = 8.8$~ in KT. Our results are consistent with
the studies of Singh \cite{Singh_PRB} which compared covalencies
in KT and KN.  In layered Nb/Ta superstructures
$Z^{\ast}_{Nb,zz}$~ {\it decreases} with the number of nn
Ta-planes, and $Z^{\ast}_{Ta,zz}$~ {\it increases} with the number
of nn Nb-planes. Both $Z^{\ast}_{B,zz}$'s approach saturation
values; $Z^{\ast}_{B,xx}$'s, however, exhibit no significant
changes. These results are similar to those of previous
computations for Nb/Al supercells in
CaTiO$_{3}$-Ca(Al$_{1/2}$Nb$_{1/2}$)O$_{3}$ where it was found
that the presense of Al decreases the Nb dynamical charge
\cite{Cockayne}.

 The effects of cation ordering on $Z^{\ast}$ can be well explained
via a model in which a layered superstructure is treated as a
stacking of slabs with different \textit{local} electronic
dielectric permittivities $\varepsilon_\infty$. We
fit calculated values to the expressions \cite{Ghosez}:

\begin{equation}
\label{eqZi} Z^{\ast}_{izz} \left( {N_j } \right) =
\frac{\varepsilon_{\infty ij} + 2}{3}g_{ij}
\end{equation}

\noindent where $g_{ij}$~ does not depend on the dielectric
permittivity. In order to compute the dielectric permittivity in
the layered structure considered we used a slightly modified
expression for the permittivity of two sequential capacitors

\begin{equation}
\label{eqepsij} \varepsilon_{\infty ij} = \frac { N_{Ta} + N_{Nb}
}{N_{Ta}/\varepsilon_{\infty Ta} +
\lambda_{ij}N_{Nb}/\varepsilon_{\infty Nb}}
\end{equation}

\noindent Here $\varepsilon_{\infty Nb}$~ and $\varepsilon_{\infty
Ta}$~ are the electronic dielectric permittivities of pure
KNbO$_3$~ and KTaO$_3$~ respectively, $N_i$~ ($i=Ta, Nb$) is the
number of Ta- or Nb-planes. This expression at $\lambda = 1$~ is
explicit for the simple cubic structure. We fixed
$\varepsilon_{\infty Ta}$~ at the experimental value 5.15 and,
correspondingly, $\lambda_{Ta,Ta}=1$. Fitting the data in Table 1
yields $\varepsilon_{\infty Nb}=6.4$, $g_{Ta,Ta}=g_{Ta,Nb}=3.7$,
$g_{Nb,Nb}=g_{Nb,Ta}=3.5$, $\lambda_{Nb,Nb}=1$, and
$\lambda_{Nb,Ta}=\lambda_{Ta,Nb}=0.97$. Deviation of the
nondiagonal element of $\lambda$ from 1, and the large values of
$g$, are connected with the difference between the Lorentz field
in the simple cubic lattice and that in the perovskite structure.
The value of the electronic dielectric permittivity in KNbO$_3$~
(6.4) is consistent with that obtained from first principles computation
for a cubic structure \cite{Krakauer} (6.6).

\section*{PMN results}

 To further investigate the effects of cation ordering on dynamical
charges in complex perovskites,
Berry phase analyses were performed for various supercells in the
perovskite based system PbMg$_{1/3}$Nb$_{2/3}$O$_3$~ (PMN);
Results are listed in Table 4.

\begin{table}[htbp]
\caption{Dynamical charges $Z^{\ast}_{ii}$~ in PMN:
n.d. are nondiagonal elements
where n.d.1 corresponds to nondiagonal elements between
[0-11] and [110] while n.d.2 is between
[0-11] and [-111]. }
\begin{tabular}
{p{30pt}p{30pt}p{30pt}p{30pt}p{30pt}p{30pt}p{30pt}p{30pt}p{30pt}p{30pt}}
 &
\multicolumn{2}{p{60pt}}{\textbf{PMN [001]}} &
\multicolumn{3}{p{90pt}}{\textbf{PMN [110]}} &
\multicolumn{4}{p{60pt}}{\textbf{PMN [111]}}  \\ \hline
{}&\textbf{[001]} & \textbf{[100]}& \textbf{[001]}&
\textbf{[-120]}&\textbf{n.d.}& \textbf{[-111]}&
\textbf{[0-11]}&\textbf{n.d.1}&\textbf{n.d.2}
\\ \hline
\textbf{Mg}   & 2.55 & 1.90 & 1.90 & 2.57 & --0.06 & 2.80 & 2.76 &
0.00 &  0.00
\\ \textbf{Nb}& 7.63 & 8.76 & 8.73 & 8.23 & 0.26 & 6.53 & 7.13 &
0.06 &  0.02
\\            & 7.63 & 8.75 & 8.69 & 8.23 & 0.24 & 6.50 & 7.13 &
0.06 & 0.00
\\ \textbf{Pb}& 3.42 & 3.80 & 3.98 & 4.07 & -0.44 & 4.31 & 3.76 &
--0.04 & --0.07
\\      \par  & 3.32 & 3.90 & 3.99 & 3.60 & 0.28 & 3.02 & 4.31 &
--0.06 & --0.03
\\      \par  & 3.38 & 3.80 & 3.98 & 3.53 & 0.29 & 4.33 & 3.78 &
--0.02 & --0.05
\end{tabular}
\label{tab2}
\end{table}

As in KTN, the dynamical charges of ions in PMN depend on nn
coordination. In particular, $Z^{\ast}_{Pb}$~ is sensitive to
chemical ordering. These results are consistent with
$Z^{\ast}_{Pb}$-computations for some other lead relaxors
\cite{Belianche}. It is interesting to notice that the
largest dynamical charge components for Pb occur in the [111]
ordered structure. The nondiagonal elements are generally small
compared with the differences between the diagonal elements,
which shows that the basis vectors are close to the eigenvectors
of the dynamical charge tensor.

\section*{Conclusions}

We have computed the minimum energy [110] barrier between
adjoining [0,0,d] Li wells in fully relaxed KLT supercells.
Our results are in good agreement with values obtained by fitting
a mean field expansion to experimental data.

We find that all ions in the supercell are displaced from their
ideal positions but among the nn ions, the oxygens closest to Li
exhibit the largest sympathetic displacements; they are displaced
in the opposite direction as Li, and move closer to one another.
These results support earlier work \cite{Glinchuk} and
computations \cite{Cohen,Tupitsin}.

First-principles computation of Li-Li pair excitations in a
[K$_{14/16}$Li$_{2/16}$]TaO$_3$~ supercell strongly supports a
previous assumption \cite{Dousenau} about the nature of
$\pi$-relaxation in KLT. Our low-energy values for excitations of
this point defect are in excellent agreement with fits to
experimental data.

Computations of the Born charges in KTN and PMN indicate that
dynamical charges are sensitive functions of nearest neighbor
coordination, which can be explained by heterogeneity in the
local electronic dielectric permittivity.

\section*{Acknowledgment}

 We thank G. Kresse for providing an LDA pseudopotential for Ta.


\begin{thebibliography}{999}

\bibitem{Resta} R. Resta, {\it J.Phys.: Condens. Matter} {\bf 12}, R107 (2000).

\bibitem{Ghosez} Ph. Ghosez, {\it Phys. Rev. B} {\bf 58}, 6224 (1998).

\bibitem{Kresse1} G. Kresse and J. Hafner, {\it Phys. Rev. B} {\bf 47}, 558
(1993).

\bibitem{Kresse2} G. Kresse and J. Furthm\"{u}ller, {\it Phys. Rev. B} {\bf 54}, 11169
(1996).

\bibitem{Singh} D. J. Singh, {\it Ferroelectrics} {\bf 164}, 143 (1991).

\bibitem{exp_KTO_a} G. Shirane, R. Newnham, and R. Pepinsky {\it Phys. Rev.} {\bf 96}, 581 (1954).

\bibitem{Postnikov}A. V. Postnikov, T. Neumann and G. Borstel, {\it
Ferroelectrics } {\bf 164}, 101 (1995).

\bibitem{Huchli}U. T. H\"ochli, K. Knorr and A. Loidl, {\it Adv. Phys. } {\bf 39},
405 (1990).

\bibitem{Toulouse} R. K. Pattnaik, J. Toulouse, and B. George {\it Phys. Rev. B }
{\bf 62}, 12820 (2000).

\bibitem{Vugmeister}B. E. Vugmeister and M. D. Glinchuk, {\it Rev. Mod. Phys.}
{\bf 62}, 993 (1990).

\bibitem{Glinchuk}V. V. Laguta, M. D. Glinchuk, I. P. Bykov, J. Rosa, L. Jastrabik,
R. S. Klein and G. E. Kugel {\it Phys. Rev. B} {\bf 52}, 7102
(1995).

\bibitem{Cohen} I. Inbar and R. Cohen {\it Ferroelectrics }{\bf 194}, 83 (1997).

\bibitem{Trepakov} S. A. Prosandeev, V. A. Trepakov, M. E. Savinov, L. Jastrabik,
and S. E. Kapphan {\it J. Phys.: Condens. Matter} {\bf 13}, 9749
(2001).

\bibitem{Kleemann} W. Kleemann, V. Sh\"onknecht, D. Sommer, and D. Rytz,
{\it Phys. Rev. Lett. }{\bf 66}, 762 (1991).

\bibitem{Dousenau} P. Doussineau, Y. Farssi, C. Frenos, A. Levelut, K. McEnaneu,
J. Toulouse, S. Ziolkiewicz, {\it Europhys. Lett. }{\bf 24}, 415
(1993).

\bibitem{Singh_PRB}D. J. Singh, Phys. Rev. B {\bf 53}, 176 (1996).

\bibitem{Cockayne} E. Cockayne, {\it J. Appl. Phys.} {\bf 90}, 1459 (2001).

\bibitem{Krakauer} C.-Z. Wang and H. Krakauer, {\it
Ferroelectrics} {\bf 194}, 97 (1997).

\bibitem{Burton} B. P. Burton, {\it Phys. Rev. B} {\bf 59}, 6087 (1999).

\bibitem{Belianche} L. Bellaiche, J. Padilla, and D. Vanderbilt,
{\it Phys. Rev. B} {\bf 59}, 1834 (1999).

\bibitem{Tupitsin} I. I. Tupitsin, A. Deineka, V. A. Trepakov, L. Jastrabik and S. E. Kapphan,
{\it Phys. Rev. B} {\bf 64} 195111 (2001).

\end{thebibliography}
 \end{document}